# Tunable Magnetic Skyrmions in Ultrathin Magnetic Nanostructures for Cellular-Level Neurostimulation

Renata Saha[†], Kai Wu[†,*], Diqing Su[‡], and Jian-Ping Wang[†,*]

[†]Department of Electrical and Computer Engineering, University of Minnesota, Minneapolis, Minnesota 55455, USA

[‡]Department of Chemical Engineering and Material Science, University of Minnesota, Minneapolis, Minnesota 55455, USA

*Corresponding author E-mails: wuxx0803@umn.edu (K. W.) and jpwang@umn.edu (J.-P. W.)

(Dated: January 4, 2019)

**Abstract**

In 2016, the Global Burden of Disease reported that neurological disorders were the principal cause of disability-adjusted life years (DALYs) and the second leading cause of deaths. Research in the last decade has pushed neuroscience to design and implement low-cost, efficient, implantable, flexible electrodes/probes and 3D arrays for neuron stimulation and sensing. Electrical arrays used in current CMOS-based technologies can be affected by the migration of cells (such as astrocytes) that attempt to seal off the electronic devices, causing increased impedance and alternations in the electric field. In this regard, magnetic nanodevices can be better candidates. A wide assortment of magnetic skyrmion-based device ideas and models have as of late been proposed featuring their potential applications. In this paper we propose a highly tunable skyrmion-based spintronic nanodevice for neuron stimulation. The effects of tunable material and magnetic properties specifically Dzyaloshinskii-Moriya interaction (DMI), perpendicular magnetic anisotropy (PMA) constant, number of skyrmions and device dimensions on stable skyrmion nucleation and smooth skyrmion dynamics in a magnetic ultra-thin film have been extensively studied. The aim of this study was to meet the standard therapeutic specifications of neuron stimulation, which is an electric field of about 10 mV/mm for a duration of 50 µs. From Faraday's Laws of Induction, skyrmion dynamics that generates an alternating magnetic flux density induces an electric field for a certain time duration in the ultra-thin film. The results of this work show that on tuning the skyrmion dynamics, the induced electric field can reach the standard value required for neurostimulation, thereby providing a strong futuristic possibility to exploit skyrmion-based spintronic nanodevices for neuron stimulation.

1. **Introduction**

In 2017, Gooch *et.al* reported [1] that the total cost for treatment of neurological disorders [2] alone pose a burden of $800 billion each year to the healthcare budget of the United States. This led to a huge surge in research for treating neurological disorders [3-9] over the last decade. However, the complexity of the neural architecture and the lack of understanding on the "connectomics" of brain disorders [10] pose a huge challenge in treating and diagnosing neurological disorders. Overcoming this require collaborative approach from multidisciplinary domains of science and engineering for the study of brain stimulation and brain mapping/sensing. In this work, we are focusing on the tunable magnetic nanodevices for neuron stimulation.

It was Kensall Wise [11] who had instigated the fabrication of the silicon Stanford arrays for electrical stimulation of neurons. Ever since then, understanding neurological disorders have witnessed new probe materials as well as new fabrication technologies to design nanoscale and application-specific electrode arrays providing high spatial and temporal sensing as well as stimulation of neurons [12-15]. To facilitate even less neuronal tissue damage, silicon probes were followed by polymer based flexible electrodes [16, 17]. Simultaneous developments of neural probes with polymer coatings [18], hydrogels [19], surface modifications [20], customized nanostructures [21], and carbon nanotubes [22, 23] continued to improve the functional interaction of the electrode sites with the surrounding neurons. Therefore, it is evident that nanoscience can revolutionize neural engineering with precise diagnostics and therapeutics for brain disorders [24]. The state-of-art electrode arrays that require special mention include the Michigan Arrays [25], the 3D-Utah Arrays [26] and Medtronic's[1] already commercialized wirelessly chargeable Deep Brain Stimulation (DBS) device for treating patients with Parkinson's Disease.

However, electrical stimulation cause large voltage artefacts [15] and trigger charge delivery to neuron-electrode interface that clusters neighboring cells, for instance astrocytes, microglial cells etc. [23]. This causes loss of electrical charge for neuron stimulation followed by neuronal tissue damage during insertion, demanding improved electrode sensitivity and lower impedance. Fabrication of a universal interface with high selectivity, sensitivity, good charge transfer characteristics, long-term chemical and recording stability has been a brooding challenge. In contrast, magnetic stimulation by spintronic nanodevices generate magnetic fields. This will not cause glial cell interference at the neuron-electrode interface. That magnetoresistive sensors have already been used for neural mapping [27, 28], gives the future possibility of putting magnetic stimulation of neuron to success with better spatial and temporal resolution.

In this paper we are proposing the use of magnetic skyrmions in ultra-thin magnetic nanostructure for neuron stimulation. Ever since the first experimental observation of skyrmions in 2009 [29], these topological structures have been believed to be of extensive use in the next generation neuromorphic computing, cutting-edge data processors and memory devices [30-32]. The stability of skyrmions are facilitated by chiral exchange interaction and broken inversion symmetry that are inherent characteristics in bulk magnetic nanostructures or ferromagnetic metal - heavy metal (FM/HM) interfaces, of which the latter is energetically preferred [31]. Skyrmions, owing to their tiny dimension and topological stability, can be driven through magnetic multilayers with ultra-low threshold current [31] causing low power consumption devices, their velocity is proportional to the electrical current density [33] making it highly tunable and furthermore, it is remarkable how their dynamics can be controlled by defects [34].

---

[1] https://www.medtronic.com/us-en/healthcare-professionals/therapies-procedures/neurological/deep-brain-stimulation.html

## 2. Mathematical Models

The schematic view of our proposed spintronic nanodevice is shown in Figure 1. An array of 12 Néel-type skyrmions, each of diameter 20 nm separated from each other by 60 nm, having an interface-induced DMI are defined in region 1. Owing to the ultra-thin dimension of the magnetic film, hedgehog or Néel-type skyrmions are energetically preferred [35-37]. Dipolar interactions between skyrmions have been reported to have negligible influence [37] and hence they are not considered. Upon passing a current through the heavy metal along +x direction, a spin polarized current at the FM/HM interface flowing along +y direction (explained later in Methods section) is induced. This spin polarized current drives the skyrmions to bypass a barrier to region 2. The barrier is defined at the center of region 1 and region 2, with the area of the two regions being equal. It should preferably be of a higher perpendicular anisotropy constant (PMA) than the surrounding thin film, *i.e.* $Ku_b > Ku_s$ for reasons explained in the Section 3.2 (see Figure 3). Due to the movement of skyrmions from region 1 to region 2, the time-varying magnetizations are observed in both region 1 and region 2. Based on Faraday's Law of Induction, alternating magnetic flux density gives an electromotive force (EMF) in region 1 and 2:

$$\oint \boldsymbol{E} \cdot d\boldsymbol{l} = -\iint \frac{\partial \boldsymbol{B}}{\partial t} \cdot d\boldsymbol{S} \quad \ldots (1)$$

Where $\boldsymbol{B}$ is the magnetic flux density, $\boldsymbol{E}$ is the induced electric field, $\boldsymbol{l}$ and $\boldsymbol{S}$ are the contour and the surface area of region 2 (or region 1), respectively. In this work, we will focus on the magnetic and electric dynamics in region 2 exclusively. Neural engineering suggests, the neuron cells if present in this region 2, can be stimulated when there is an electric field higher than 10 mV/mm lasting for duration of about 50 µs [38-41]. This can be achieved by the gradually moving the skyrmions to region 2. For desired device performance, specific tuning can be obtained by varying the dimensions of the spintronic devices and/or barriers in the path of movement [42], the number of skyrmions [30], the applied electrical current density [30, 32, 33], introducing high anisotropy barrier acting as defects [30,42] in the magnetic thin film and tuning the DMI until a stable skyrmion is nucleated [37,42].

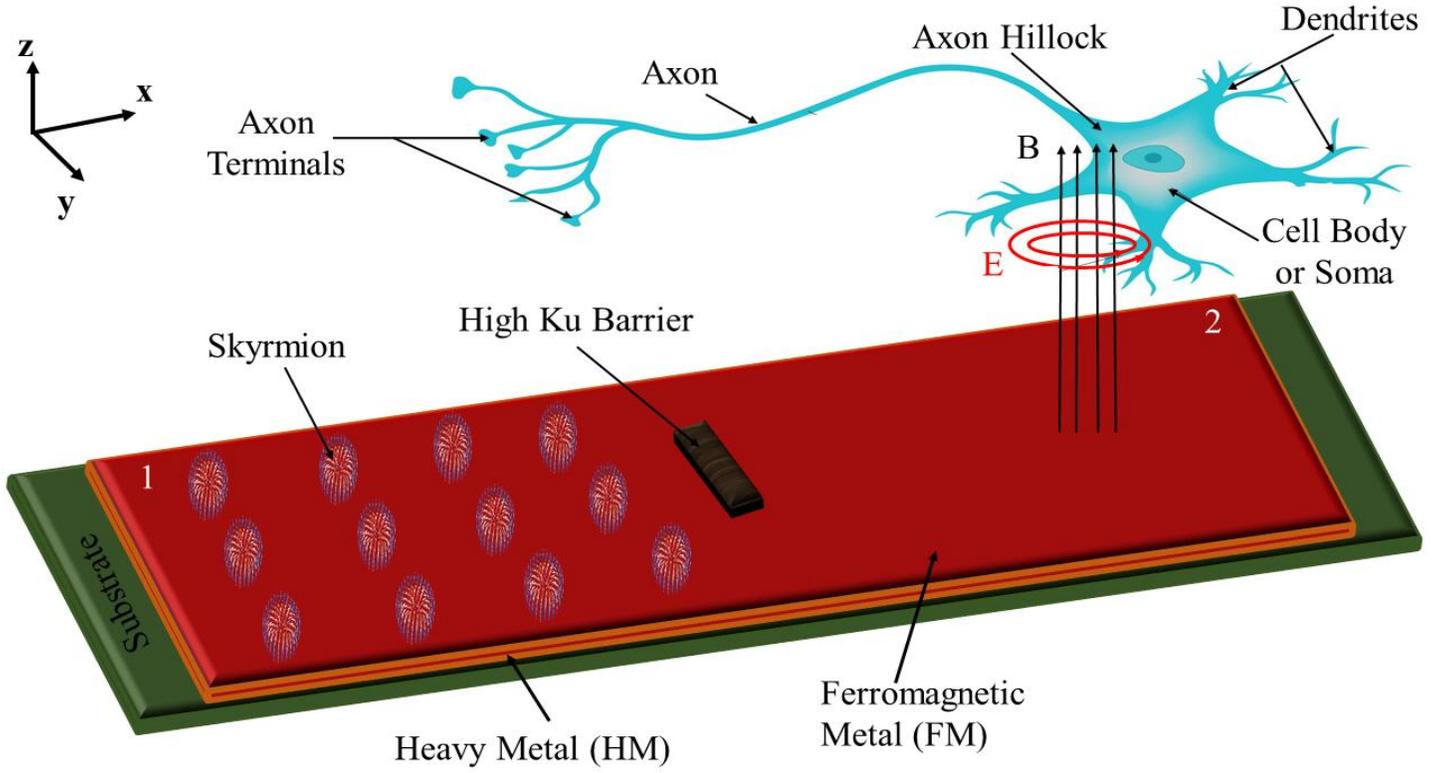

Figure 1. Schematic view of the proposed tunable skyrmion-based spintronic device with a high Ku barrier at its center. Passing electric current through the heavy metal (HM) could generate transverse spin current (flows along +y direction) that drives the array of skyrmions through the ferromagnetic metal (FM) thin film from region 1 to region 2 (called Spin Hall Effect (SHE) across HM/FM interface). This phenomenon causes a change of magnetization $M$ and an alternating magnetic flux density $B$ in region 2. According to Faraday's Law of Induction, the induced electric field $E$ at region 2 can stimulate neurons.

The proposed tunable skyrmion-based spintronic device has been numerically simulated based on the Landau–Lifshitz–Gilbert (LLG) equation with a spin transfer torque (STT) term as follows [30]:

$$\frac{d\boldsymbol{m}}{dt} = \gamma_0 \boldsymbol{h}_{eff} \times \boldsymbol{m} + \alpha \boldsymbol{m} \times \frac{d\boldsymbol{m}}{dt} + \frac{u}{t}\boldsymbol{m} \times (\boldsymbol{m}_p \times \boldsymbol{m}) \quad \ldots (2)$$

Where, $\boldsymbol{m} = \boldsymbol{M}/M_s$ is the normalized magnetization, $\gamma_0 = -2.211 \times 10^5$ mA$^{-1}$s$^{-1}$ is the gyromagnetic ratio, $\boldsymbol{h}_{eff} = \boldsymbol{H}_{eff}/M_s$ is the reduced effective field, $t$ is the thickness of the ferromagnetic (FM) layer, $\boldsymbol{m}_p$ is the current polarization vector, $u = \gamma_0(\frac{\hbar j P}{2eM_s})$, and $j$ is the density of the spin current. The direction of the spin polarization (P) is dependent on the spin hall angle of HM layer and the electric current direction. In this work, we defined P along the +y direction. The values of parameters $M_s$, $P$, $\alpha$ along with dimensions of the magnetic thin film are listed in Table 1. All parameters are adopted from Ref.[43], for $t = 0.4$ nm thick Co (FM) layer on Pt (HM) layer.

The skyrmion dynamics can be explained by the Thiele Equation [42]:

$$\mathbf{G} \times (\mathbf{v}^s - \mathbf{v}^d) - D\alpha \mathbf{v}^d - \mathbf{F}(x) = 0 \quad \ldots (3)$$

Where, $D$ is the magnitude of DMI, $\mathbf{v}^d$ is the drift velocity of the skyrmions, $\mathbf{v}^s$ is the velocity induced by spin current. $\mathbf{G} = (0,0,4\pi Q)$ is the gyromagnetic coupling vector representing the Magnus force, Q is the topological winding number of the skyrmion given by, $Q = -\int dx \frac{1}{4\pi} \mathbf{m}(x) \cdot (\partial_x \mathbf{m}(x) \times \partial_y \mathbf{m}(x))$, and $F(x) = \nabla V(x)$ is the force induced by skyrmion potential. The potential is given by the local energy of a skyrmion [42]:

$$E_{sk} = -\frac{D^2 \pi^4}{4K\pi + \frac{16}{\pi}B} + 38.7A \quad \ldots (4)$$

Where, $B$ is the flux density (here, $B = \mu_0 M$), $A$ is the exchange constant, $K$ is the anisotropic energy of the film (here, $Ku_b$ and $Ku_s$, depending on the region of the presence of the skyrmion). Also, radius of the nucleated skyrmion is given by [42]:

$$R_{sk} = -\frac{D\pi^2}{2K\pi + \frac{8}{\pi}B} \quad \ldots (5)$$

The values of remaining parameters such as $A$ and $\mu_0$ are listed in Table 1. The tuning of these parameters to suffice our application will be discussed in the next section. The numerical and computational assumptions have been explained in the Methods section.

Table 1. Simulation parameters for movement of skyrmion along a magnetic thin film.

| Parameters | Description | Values |
|---|---|---|
| **Thin Film Dimension** | Length × Width × Thickness | 500 nm × 200 nm × 1 nm |
| **Cell Size** | Length × Width × Thickness | 2 nm × 2 nm × 1 nm |
| $\alpha$ | Gilbert damping factor | 0.3 |
| A | Exchange constant | $15 \times 10^{-12}$ J/m |
| $\lambda$ | Slonczewski $\Lambda$ parameter | 1 |
| $\varepsilon$ | Slonczewski secondary STT term | 0 |
| P | Polarization factor | 0.4 |
| $M_s$ | Saturation magnetization | $580 \times 10^3$ A/m |
| J | Electrical current density | $1 \times 10^{10}$ A/m$^2$ |
| $\mu_0$ | Permeability of free space | $4\pi \times 10^{-7}$ WbA$^{-1}$m$^{-1}$ |

### 3. Results and Discussion

*3.1 Current driven movement of skymions and the induced electromotive force (EMF)*

Upon applying an electric current density $J = 1 \times 10^{10}$ Am$^{-2}$ in the heavy metal (HM), the spin polarized current driven motion of the nucleated skyrmions have been observed in presence of a barrier with higher Ku than the surroundings. Figure 2a gives the simulated snapshots at a time step of 10 ns showing the skyrmions in motion, nucleated with DMI = 3 mJ m$^{-2}$. The skyrmions move past the barrier with dimensions, $y_b \times x_b$ = 40 nm × 40 nm and PMA constant, $Ku_b$ = 0.84 MJ m$^{-3}$, when the surrounding thin film has $Ku_s$ = 0.7 MJ m$^{-3}$. The normalized magnetization ($m_z$) in region 2 changes over a time duration of ~80 ns (see Figure 2b), so does the induced E-field in region 2.

From Figure 2b, $\Delta m_z = 0.2, \Delta t = 80$ ns, thus the induced electric field in region 2 for this configuration is calculated to be $E = 0.1$ mV/mm with a duration of 80 ns (from equation (1)). However, the required field for neuron stimulation is 10 mV/mm with a duration of about 50 μs [38]. Therefore, it is evident that we need to further impede the skyrmions to increase the time duration $\Delta t$ to the orders of several microseconds. On a different note, the intensity of the induced electric field is also insufficient to stimulate a neuron. However, as a direct consequence of the Faraday's Law of Induction, from equation (1), $E$ decreases as $\Delta t$ increases. Therefore, to increase $E$, another option is to increase $\Delta m_z$ in region 2. Also, $E$ will increase if $\Delta S \gg \Delta l$. For the present rectangular device shape, increasing the device dimension will automatically lead to $\Delta S \gg \Delta l$. In summary, for neuron stimulation there must be a simultaneous increase in $\Delta m_z$ and $\Delta t$. Rigorous tuning of present material and magnetic parameters of the spintronic device is essential to enable a suitable device performance for neuron stimulation.

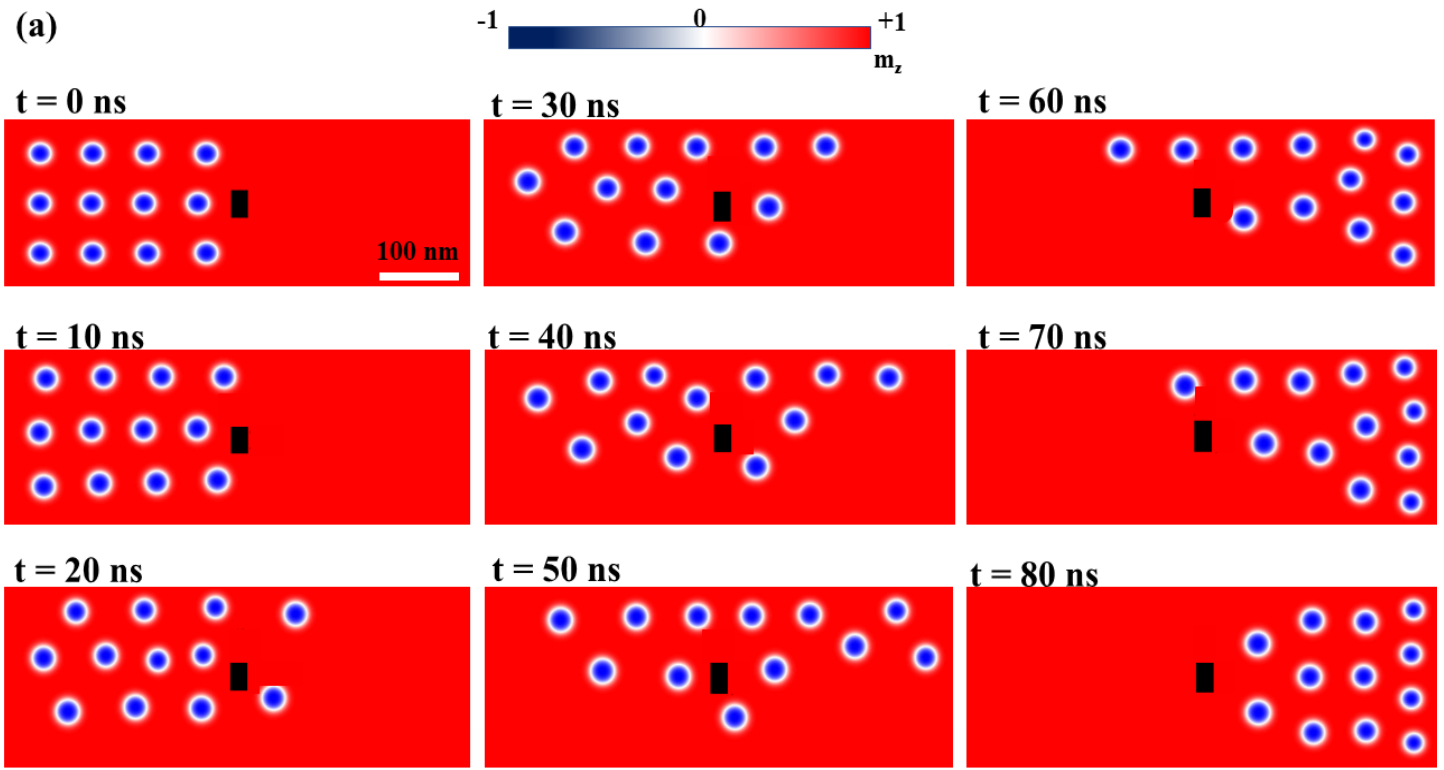

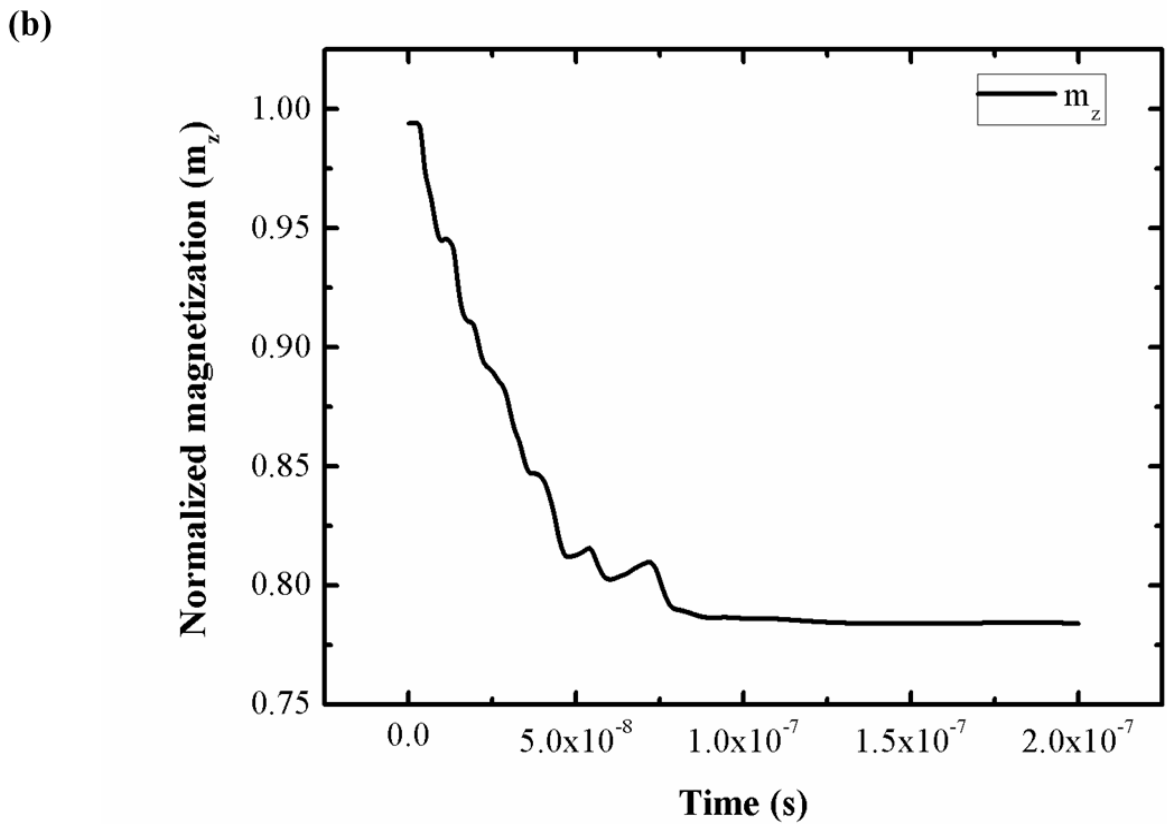

Figure 2. (a) Micromagnetic simulation snapshots at a time step of 10 ns for an array of 12 skyrmions moving across a FM/HM thin film of dimensions 500 nm × 200 nm × 1 nm, DMI = 3 mJ m$^{-2}$, Ku$_s$ = 0.7 MJ m$^{-3}$, electric current density $J = 1 \times 10^{10}$ A m$^{-2}$, the barrier has dimensions 40 nm × 40 nm and Ku$_b$ = 0.84 MJ m$^{-3}$.

The magnetic skyrmions bypass the barrier from region 1 to region 2. The skyrmionic movement in magnetic thin film can be found in Supplementary Movie S1. (b) The time-resolved normalized magnetization (the average magnetization component along the z direction) in region 2 of the magnetic thin film.

*3.2 Tuning the combinations of DMI and $Ku_s/Ku_b$ ratio*

Figure 3 shows the magnetization patterns of the magnetic thin films with identical barrier dimensions, $y_b \times x_b = 80$ nm $\times$ 40 nm at run-time, $\tau = 200$ ns but different combinations of DMI and $Ku_s/Ku_b$ ratios. These different combinations of magnetic parameters are tuned to study their effect on the stable nucleation of skyrmions and their smooth movement through the thin film into region 2. Experimentally, magnetic anisotropy can be tuned by ion irradiation and DMI strength can be reduced by increasing the film thickness [37].

Three different characteristics have been observed: (i) The diagonal of the working window in Figure 3 shows the most promising performance. The cases for DMI = 2.5 mJ m$^{-2}$, $Ku_s/Ku_b = 0.8$ and for DMI = 3 mJ m$^{-2}$, $Ku_s/Ku_b = 1$ satisfy both requirements for stable nucleation of skyrmions and the smooth movement bypass the barrier. (ii) For the cases below the diagonal, the DMI values are small. Hence, the skyrmions nucleate with smaller radii which agrees with equation (5). For the case where the DMI = 2 mJ m$^{-2}$, $Ku_s/Ku_b = 0.8$, the energy from equation (4) is just sufficient to nucleate skyrmions and it takes a longer time to stabilize. This leads to the annihilation of skyrmions immediately after their nucleation (see Supplementary Movie S2). For the remaining cases below the diagonal, the combinations of D and $Ku_s/Ku_b$ values lead to low energy of the skyrmions which is insufficient for the skyrmion nucleation (see equation (4)). (iii) For cases above the diagonal, the large DMI values increase both the energy as well as the radius of the skyrmions above the threshold, leading to the formation of Néel stripes [44]. These elongated Néel stripes form fast and spread all over the magnetic ultra-thin, causing a huge change of magnetization throughout the entire thin film. Yet they remain static at their position of nucleation, even after passing electric current, which implies there is little change in time duration. The video for the case of DMI = 4 mJ m$^{-2}$, $Ku_s/Ku_b = 0.6$ for $\tau = 200$ ns is attached in Supplementary Movie S3. Thus, a large change in magnetization takes place over a very small duration of time, which is not applicable for neuron stimulation.

In summary, Figure 3 gives us the combinations of the values DMI = 3 mJ m$^{-2}$, $Ku_s/Ku_b = 1$ and DMI = 2.5 mJ m$^{-2}$, $Ku_s/Ku_b = 0.8$ that are energetically favorable for stable skyrmion nucleation in region 1 and smooth movement bypassing the barrier to region 2. This working window also confirms a previous statement that the PMA of the barrier should preferably be greater than that of the surroundings, $Ku_b \geq Ku_s$. For some of the cases, it is seen that stable skyrmions are generated and some of them move to the region 2 with only a few remaining in region 1; this phenomenon can be justified by the effects of $Ku_s/Ku_b$ ratio in association with the barrier dimensions which will be discussed in Section 3.3. Another thing to be noted is that, the combination

DMI and $Ku_s/Ku_b$ ratio does not alter the time change of magnetization by large folds. More precisely, tuning this combination controls the nucleation of skyrmions, not it's dynamics. For all cases in Figure 3, where there is stable nucleation and smooth movement of the skyrmions, the time duration for change in magnetization value in region 2 is nearly the same, which is ~100 ns.

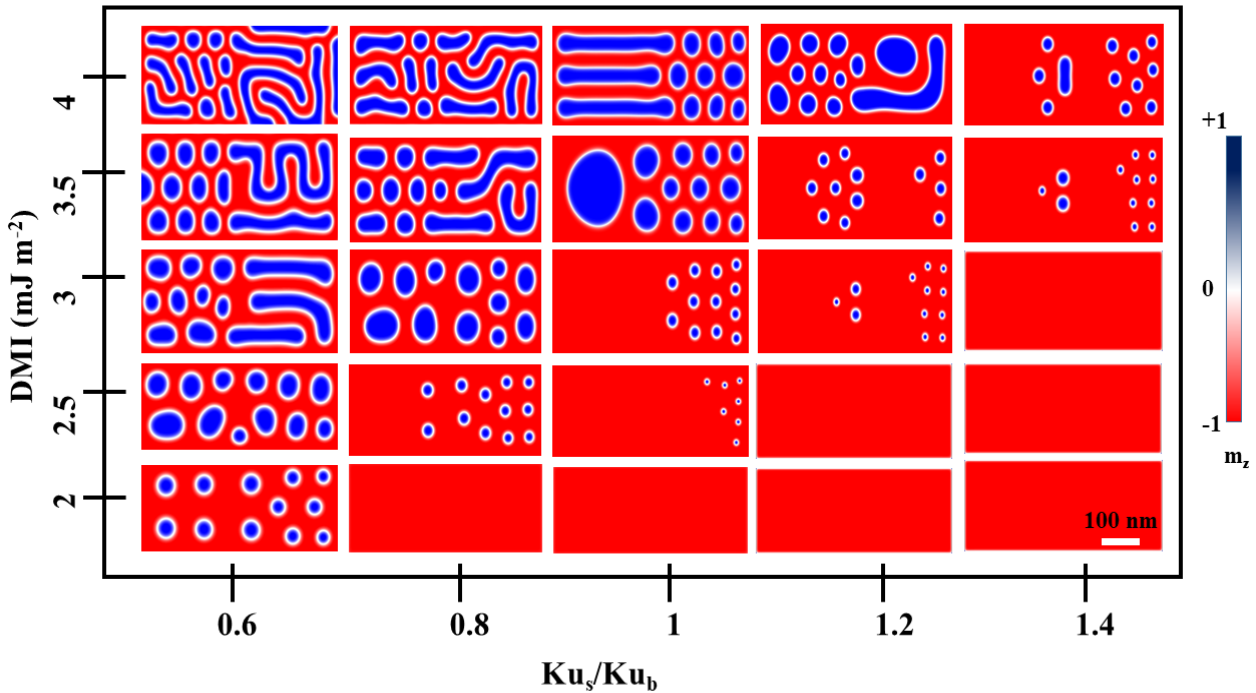

Figure 3. At τ = 200 ns, the working window of the movement of skyrmions to region 2 under different combinations of DMI and $Ku_s/Ku_b$, the barrier dimension is identical $y_b \times x_b$ = 80nm × 40nm. Our requirements for neuron stimulation are best satisfied along the diagonal of the window, the most promising device performance being the cases where DMI = 2.5 mJ m$^{-2}$, $Ku_s/Ku_b$ = 0.8 and DMI = 3 mJ m$^{-2}$, $Ku_s/Ku_b$ = 1. This confirms that $Ku_b$ must preferably be greater than or equal to $Ku_s$. Below the diagonal in that window, the skyrmions are either nucleated in smaller diameters (for cases where DMI = 2.5 mJ m$^{-2}$, $Ku_s/Ku_b$ = 1; DMI = 3 mJ m$^{-2}$, $Ku_s/Ku_b$ = 1.2 and DMI = 3.5 mJ m$^{-2}$, $Ku_s/Ku_b$ = 1.4) or annihilate immediately after their nucleation (for DMI = 2 mJ m$^{-2}$, $Ku_s/Ku_b$ = 0.8, see Supplementary Movie S2) or like the remaining cases below the diagonal where skyrmions never nucleate. For cases above the diagonal, they form Néel stripes that are not useful for our application in neuron stimulation (for DMI = 4 mJ m$^{-2}$, $Ku_s/Ku_b$ = 0.6, see Supplementary Movie S3).

*3.3 Tuning the combination of $Ku_b/Ku_s$ and $y_b/x_b$ ratios*

The working window in Figure 4 demonstrates the snapshots of the magnetic thin film with a desired DMI = 3 mJ m$^{-2}$ at τ = 200 ns with varying PMA ratios of the barrier to the surroundings and varying length to width ratios of the barrier in the nanodevice.

The results of the working window are concluded as followings: (i) For $Ku_b = Ku_s$, under any $y_b/x_b$ ratio, the skyrmions face no barrier in their path of movement. However, although all skyrmions reach to region 2, the time duration $\Delta t$ is short (i.e., the time window where magnetization keeps increasing/decreasing and E-field is non-zero). As a result, there will be insufficient electric field duration for neuron stimulation. (ii) For $Ku_b < Ku_s$, the barrier acts as a sink. While most of the skyrmions bypass the barrier to region 2, a few skyrmions are trapped in the low Ku barrier. In this case, when $y_b > x_b$, the skyrmions that were trapped in the low Ku barrier act as a barrier in the path of the remaining skyrmions. This also explains why few skyrmions for some of the cases below the diagonal of the working window in Figure 3 did not reach to region 2. In this case, as not all skyrmions arrive at region 2, $\Delta m_z$ will be small and unable to induce sufficient electric field for neuron stimulation. (iii) For $Ku_b > Ku_s$, the barrier acts as a hurdle in the path of skyrmion movement. Thus, the skyrmions avoid the barrier on their way to region 2. As $y_b$ becomes larger than $x_b$, within a time window of 200 ns, it is observed that fewer skyrmions arrive at region 2. Also, since $Ku_b > Ku_s$, the barrier will not act as a sink to trap the skyrmions. It is expected that one can obtain larger $\Delta m_z$ and $\Delta t$ simultaneously as more skyrmions are likely to move into region 2 after τ = 200 ns. This meets our requirements for neuron stimulation as was discussed in Section 3.1. In the highlighted dashed square in Figure 4, the cases where $Ku_b/Ku_s = 1.2, y_b/x_b = 1.4$ and $Ku_b/Ku_s = 1.4, y_b/x_b = 1.4$ give us the most promising results for our proposed spintronic nanodevice for neuron stimulation and the change in magnetization in the region 2 lasts for ~110 ns and ~140 ns, respectively.

It is worth to mention that, $y_b > x_b$ affects the dynamics of skyrmions whereas for cases $y_b < x_b$ does not affect the dynamics. This is because the alignments of the array of skyrmions are different in region 1 (see Figure 2(a)) and in region 2. Also, for a barrier with larger PMA values, the skyrmions do not pass through the barrier, instead, they pass around the barrier. Hence, the y dimension of the barrier affects their dynamics more than its x dimension.

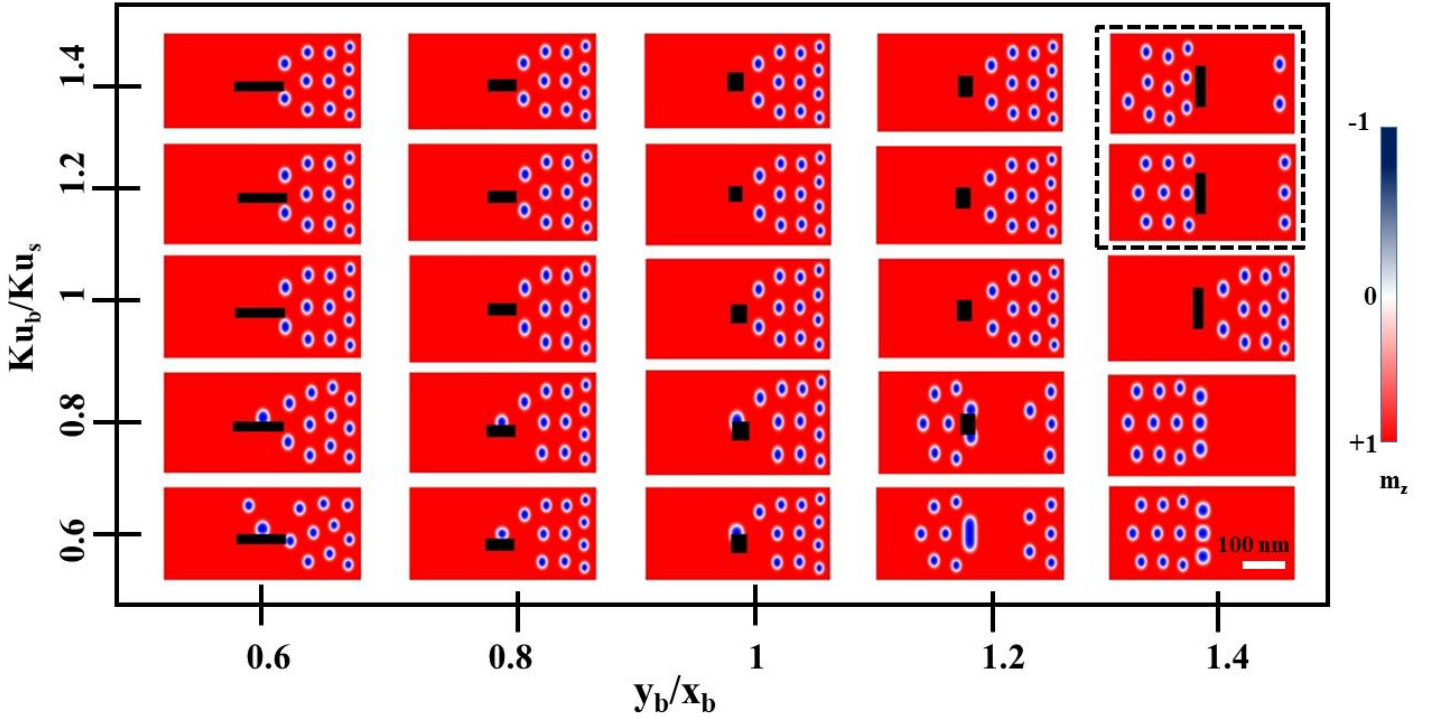

Figure 4. At $\tau = 200$ ns, the working window of the movement of skyrmions to region 2 under different combinations of $Ku_s/Ku_b$ and $y_b/x_b$ of the nanodevice. (i) When $Ku_b = Ku_s$, the skyrmions face no barrier in their path of movement. Although all skyrmions arrive at region 2, the change in magnetization will not last long enough to successfully stimulate a neuron, $i.e. \Delta t\ is\ small$. (ii) When $Ku_b < Ku_s$, the barrier acts as a sink. While most of the skyrmions bypass the barrier to region 2, a few skyrmions are trapped in the barrier. In this case, when $y_b > x_b$, the skyrmions that were trapped in the low Ku barrier act as a barrier in the path of the remaining skyrmions and some skyrmions do not reach region 2 at all. For this case in region 2, $\Delta m_z\ is\ small$. (iii) When $Ku_b > Ku_s$, the barrier acts as a hurdle in the path of skyrmion movement. As $y_b$ becomes larger than $x_b$, within a run-time of 200 ns, it is observed that fewer skyrmions pass, which implies that by increasing the number of skyrmions and run-time $\tau > 200$ ns, it is possible to obtain large change in magnetization over a larger time duration, $i.e.$ increased $\Delta m_z$ and increased $\Delta t$, simultaneously. The cases highlighted in dashed square are expected to give the most satisfactory results for neuron stimulation.

*3.4 Tuning the number of nucleated skyrmions*

The material and magnetic properties of the highlighted cases of Figure 4 satisfies our requirements for neuron stimulation, thereby promising a simultaneous increase in $\Delta m_z$ and $\Delta t$. By setting $Ku_b/Ku_s = 1.2$ and $y_b/x_b = 1.4$ (one of the two highlighted cases from Figure 4) and the run-time of 200 ns, nanodevices with 12 skyrmions and the 16 skyrmions were simulated. Figure 5 shows that for the nanodevice with 12 nucleated skyrmions, $\Delta m_z = 0.06, \Delta t = 110$ ns and the calculated induced electric field for neuron stimulation is $0.022\ mV/mm$ for a time duration of 110 ns. For the other nanodevice with 16 nucleated skyrmions, $\Delta m_z = 0.15, \Delta t = 170\ ns$ and the calculated induced electric field for neuron stimulation is $0.0357\ mV/mm$ for a duration of 170 ns. Thus, region 2 of the nanodevice with 16 nucleated skyrmions show larger $\Delta m_z$ and $\Delta t$ in comparison to the device with 12 nucleated skyrmions. Therefore, by keeping the dimensions of the device constant and meanwhile increasing the skyrmion numbers, we can achieve larger electric field and time duration in region 2 for neuron stimulation.

Therefore, to magnetically stimulate neuron cells, if the proposed tunable skyrmionic nanodevice dimensions can be increased to the order of micrometers, then for the rectangular region 2, $\Delta S \gg \Delta l$. If this is followed by a simultaneous increase in the number of skyrmions in region 1, the barrier dimension ratio being in the order $y_b/x_b = 1.4$ and the PMA ratio being in the order of $Ku_b/Ku_s = 1.2$, the induced electric field intensity ($E$) in the region 2 can be conclusively predicted to increase as (i) $\Delta S$ will increase largely due to increase in device dimension; (ii) $\Delta m_z$ in the region 2 will increase due to arrival of larger number of skyrmions; (iii) $\Delta t$ for the process will increase as an increased number of skyrmions have to traverse through an increased area of the spintronic device. Hence, from the above analysis we can conclude that a magnified version of this proposed skyrmion-based spintronic device is applicable for neuron stimulation. However, in this work, we have limited our study to devices with several hundred nanometers size considering the simulation time.

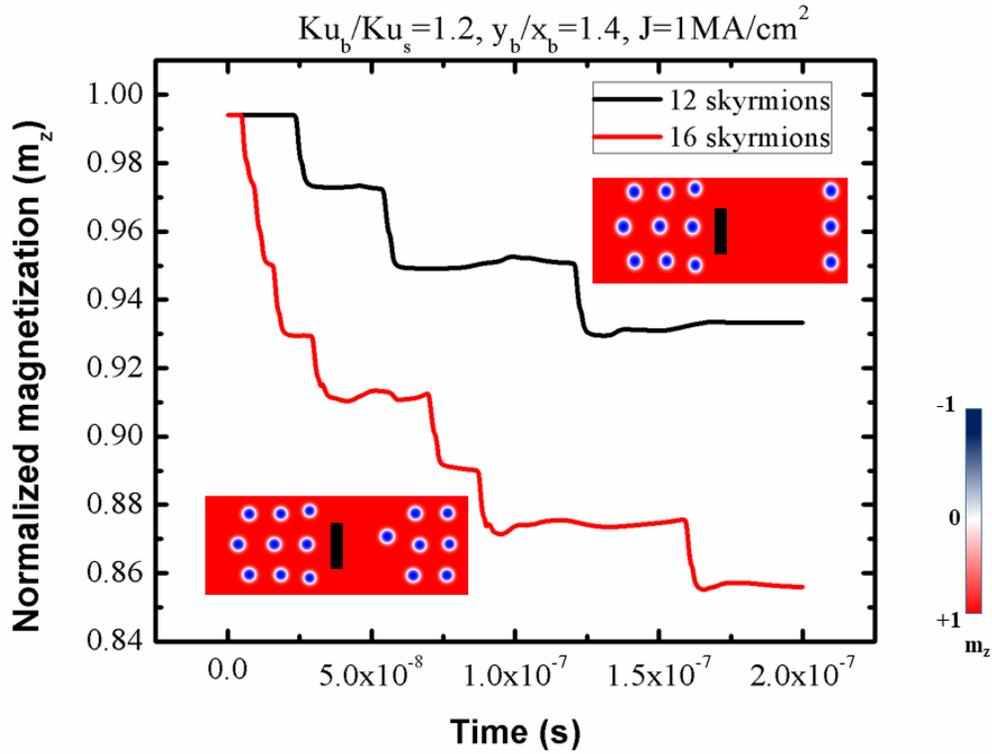

Figure 5. For $Ku_b/Ku_s = 1.2$ and $y_b/x_b = 1.4$, two cases are investigated within a run-time of 200 ns: (i) for the case of 12 skyrmions, the rate of change of magnetization in region 2 is ~110 ns. (ii) For the case of 16 skyrmions, the rate of change of magnetization in region 2 is ~170 ns. This implies that if we can increase the number of the skyrmions in a device size of the orders of a few micrometers, keeping the skyrmion size and gap identical, we can increase the time duration to the orders of several microseconds as well as the change of magnetization to achieve the induced electric field magnitude for neuron stimulation.

4. Conclusion

Our objective was to assess realistically the scope of spintronics for neuron stimulation in future. To conclude, we have proposed a highly tunable magnetic skyrmion-based spintronic nanodevice for this purpose. We have used a barrier with a high PMA constant in a magnetic ultra-thin film to meet certain therapeutic specifications for our application. Also, we have studied the tuning of the combinations of DMI and PMA ratios of the surrounding to the barrier on stable nucleation of skyrmions in a spintronic device. It is observed that the dynamics of the skyrmions under a certain electric current density are affected by the PMA and the dimension of the barrier. Finally, it is expected that by increasing skyrmion number in the magnetic ultra-thin film, it is possible to meet all the requirements for single neuron stimulation. The nucleation, stabilization and manipulation of a group of skyrmions in and along defects is not a breakthrough as they have been analytically predicted earlier [30, 37]. But the perfect controlling of those nucleated skyrmions in a magnetic thin film such that they can meet the specifications of a neuron stimulating device has been the primary focus of this work. Besides, all the present implantable neurostimulation devices are have their dimensions in the order of cm or mm (see Table 1 of [15]). The highly scalable property of spintronic devices (in this work in the orders of a few micrometers), low power and flexibility show their bright future in neural engineering. Apart from skyrmions, the possibility of other dynamic spintronic nanostructures such as domain walls in nanowires, vortices can also be studied for the same purpose.

**Methods**

Mumax3 along with Dzyaloshinskii-Moriya interaction (DMI) module was used to perform the three-dimensional micromagnetic simulations. The time dependent magnetization dynamics is determined by the Landau-Lifshitz-Gilbert Equation (LLG) equation including the spin transfer torque (STT) term. The average energy density calculations include the exchange energy, the anisotropy energy, the applied field (Zeeman) energy, the magnetostatic (demagnetization) energy and the DMI energy terms.

In all the simulations in this work, the dimension of the magnetic thin film is set as 500 nm × 200 nm × 1 nm. All models are discretized into cells with sizes of 2 nm × 2 nm × 1 nm in the simulation which is sufficiently smaller the skyrmion diameter (20 nm) which guarantees numerical precision along with computational efficiency. The run time of the simulation is fixed at 200 ns. Other parameters used in this work are listed in Table 1.

The initial magnetic states of the thin film are relaxed along +z direction, except for the tilted magnetization near the edges due to the DMI. Initially, an array of 12 skyrmions are nucleated in region 1 of the magnetic thin film at desired positions, 60 nm apart from each other as shown in Figure 1. Then they are moved from region 1 to region 2 by the transverse spin-polarized current and then relaxed to stable/metastable state within a short period of time. For simplicity, we directly simulate the injection of spin-polarized current with certain current density and polarization rate.

**Supplementary Materials**

Supplementary movie S1. The movement of 12 skyrmions, nucleated with DMI $= 3$ mJ m$^{-2}$, moving through a magnetic thin film with Ku$_s = 0.7$ MJ m$^{-3}$, for an applied current density $J = 1 \times 10^{10}$ A m$^{-2}$ bypassing a barrier having dimensions $y_b \times x_b = 80$ nm × 40 nm and Ku$_b = 0.84$ MJ m$^{-3}$, for τ = 200 ns.

Supplementary movie S2. The annihilation of skyrmions due to nucleation with insufficient DMI $= 2$ mJ m$^{-2}$, moving through a magnetic thin film with a centered barrier, having Ku$_s$/Ku$_b = 0.8$, for an applied current density $J = 1 \times 10^{10}$ A m$^{-2}$, for τ = 200 ns.

Supplementary movie S3. The formation of Néel stripes due to nucleation with large DMI $= 4$ mJ m$^{-2}$, moving through a magnetic thin film with a centered barrier, having Ku$_s$/Ku$_b = 0.6$, for an applied current density $J = 1 \times 10^{10}$ A m$^{-2}$, for τ = 200 ns.

**Acknowledgements**

This study was financially supported by the Institute of Engineering in Medicine of the University of Minnesota, National Science Foundation MRSEC facility program, the Distinguished McKnight University Professorship, Centennial Chair Professorship, Robert F Hartmann Endowed Chair, and UROP program from the University of Minnesota.

**Conflict of interest**

The authors declare no conflict of interest.